\documentclass[conference,10pt]{IEEEtran}
 \IEEEoverridecommandlockouts
\usepackage{amssymb}
\usepackage{amsmath}

\usepackage{cite}

   \usepackage[dvips]{graphicx}

   \DeclareGraphicsExtensions{.eps,.pdf}

\ifCLASSOPTIONcompsoc
\usepackage[tight,normalsize,sf,SF]{subfigure}
\else
\usepackage[tight,footnotesize]{subfigure}
\fi

\begin{document}
%
\title{Joint Uplink and Downlink Relay Selection in Cooperative Cellular
Networks \thanks{This paper is supported by the National Science
Foundation of China (NSFC 60702051, NSFC-AF: 60910160), the
Specialized Research Fund for the Doctoral Program of Higher
Education (SRFDP：20070013028), and the Program for New Century
Excellent Talents in University (NCET-08-0735). This paper is
co-funded by Nokia on the Beyond 3G Research project.}}

%
%

\author{\IEEEauthorblockN{Wei Yang\IEEEauthorrefmark{1}, Lihua Li\IEEEauthorrefmark{1}, Gang Wu\IEEEauthorrefmark{2}, Haifeng Wang\IEEEauthorrefmark{2}, and Ying Wang\IEEEauthorrefmark{1}}
\IEEEauthorblockA{\IEEEauthorrefmark{1}Key Lab. of Universal
Wireless Commun., Beijing University of Posts and Telecom.(BUPT),\\
Ministry of Education, Wireless Technology Innovation
Institute(WTI), BUPT, China}
\IEEEauthorblockA{\IEEEauthorrefmark{2}Wireless Modem System
Research, Device R\&D, NOKIA, ShangHai, P.~R.~China} }

%
%


%



\maketitle

\begin{abstract}
We consider relay selection technique in a cooperative cellular
network where user terminals act as mobile relays to help the
communications between base station (BS) and mobile station (MS). A
novel relay selection scheme, called Joint Uplink and Downlink Relay
Selection (JUDRS), is proposed in this paper. Specifically, we
generalize JUDRS in two key aspects: (i) relay is selected jointly
for uplink and downlink, so that the relay selection overhead can be
reduced, and (ii) we consider to minimize the weighted total energy
consumption of MS, relay and BS by taking into account channel
quality and traffic load condition of uplink and downlink.
Information theoretic analysis of the diversity-multiplexing
tradeoff demonstrates that the proposed scheme achieves full spatial
diversity in the quantity of cooperating terminals in this network.
And numerical results are provided to further confirm a significant
energy efficiency gain of the proposed algorithm comparing to the
previous best worse channel selection and best harmonic mean
selection algorithms.

\end{abstract}

\begin{IEEEkeywords}
relay selection, cooperative networks, energy-efficient, asymmetric
traffic, weighted energy consumption.
\end{IEEEkeywords}

%
\IEEEpeerreviewmaketitle

\section{Introduction}
Cooperative relaying is a promising technology that can not only
increase the overall throughput and energy efficiency, but also
enable the system to guarantee the quality of service (QoS) desired
by the various media classes for the next-generation wireless
communications\cite{Laneman}.
\par
However, the relay nodes consume system resources and energy, hence
limiting the transmission rate and energy efficiency. As a result,
various relay selection schemes have been introduced in previous
works. Nosratinia and Hunter\cite{Gouping} demonstrate that relay
selection techniques can capture maximum diversity in the number of
cooperating nodes, while each node only knows its own receive
channel state. Madan et al.\cite{patent} consider selecting relays
by minimizing the total power consumption. Tannious et
al.\cite{ITRS} propose an ITRS protocol, which employs hybrid-ARQ
with packet combining at the destination and includes a
limited-feedback handshake for relay selection, achieving the
multiple-input single-output (MISO) diversity-multiplexing tradeoff
(DMT) bound. Bletsas et al.\cite{opportunistic} propose a
decentralized, opportunistic relaying scheme to select the best
relay based on instantaneous end-to-end channel conditions, which
also achieves the MISO DMT bounds.
\begin{figure}
\centering
\includegraphics[width=2.7in]{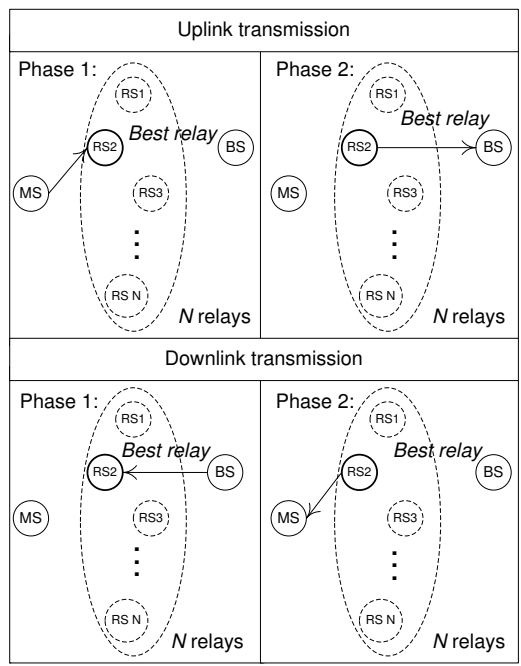}
\caption{We consider communication between BS and MS via one mobile
relay. In the uplink transmission, the relay forwards the signal
from MS to BS, and in downlink, the same relay forwards the signal
from BS to MS.} \label{moto}
\end{figure}
\par
The above-mentioned works all focus on single direction
transmission, from source to destination. However, in most cellular
networks, the uplink and downlink may be in deep fading at the same
time. In this scenario, relay is needed for both uplink and
downlink, and the same relay can be adopted if the channel
reciprocity exists. Base station (BS) and mobile station (MS) can
act interchangeably as source and destination, with the relay
offering help for both two communications (see Fig.~\ref{moto}). In
this paper, we consider to select the best relay for uplink and
downlink jointly in one selection, which significantly reduce the
relay selection overhead.
\par
Meanwhile, the traffic load of uplink and downlink have been
changing to be asymmetric to support new multimedia communication
services. For example, the downlink traffic load may be larger than
that in the uplink when MS downloads some files from BS. In this
scenario, since the relay transmits signal to MS more often than to
BS, the MS-relay channel and BS-relay channel may have different
priority on relay selection. Motivated by this, we take into account
the traffic load condition of uplink and downlink during the
selection of relay.
\par
%
\par
This paper is organized as follows. In the next section, we present
the syetem model. In Section \ref{JUDRS}, we describe the JUDRS
protocol. Diversity-multiplexing tradeoff performance of the
proposed scheme is analyzed in Section \ref{DMT}, and numerical
results of energy-efficiency of JUDRS are discussed in Section
\ref{numerical results}. Finally, conclusions are drawn in Section
\ref{conclusion}.

\section{System Model}
\label{system_model} We consider a half-duplex dual-hop
communication scenario in a fading environment with one MS, one BS
and a set $S=\{1,2,\cdots N\}$ of $N$ decode-and-forward (DF)
relays, depicted in Fig.~\ref{moto}.
During the first hop, the source transmits its information to the
relays and destination, while in the second phase, one relay (assume
that each time, at most one relay is used) decodes and forwards the
received signal using the same modulation and coding scheme (MCS).
The destination receiver combines the messages it receives from the
source and relay using optimal diversity combining. In cooperative
cellular networks, MS and BS act interchangeably as source and
destination, in the UL and DL transmission, with relay offering help
for both the UL and DL communications.

\par
The channels from MS to relays and from relays to BS are frequency
non-selective channels that undergo independent Rayleigh fading.
Thus, the channel gains from MS to relays, denoted by $|h_i|^2$, and
from relays to BS, denoted by $|g_i|^2$, are independent and
exponentially distributed random variables with
parameter $\lambda_{M,i}$ and $\lambda_{B,i}$, respectively, where
$i=1,\cdots N$. The mean channel gains depend on shadowing and the
distance between corresponding nodes, thus we get,
\begin{equation}
\frac{1}{\lambda_{i,j}}=\left(\frac{\sqrt{G_TG_R}\mu}{4\pi
d_0}\right)^2\cdot \left(\frac{d_0}{d_{i,j}}\right)^p\cdot
\psi_{i,j},
\end{equation}
where $\mu$ is the carrier wavelength, $G_T$ and $G_R$ are the
transmitter and receiver antenna gain respectively, $d_0$ is a
reference distance for the antenna far field, $p$ is the path loss
exponent, and $ d_{i,j}$ and $\psi_{i,j}$ are the distance and
shadow fading between corresponding nodes. We assume that the UL and
DL of each link are reciprocal. This condition is fulfilled in TDD
systems where round-trip duplex time is much shorter than coherence
time of the channel.
\par
At all nodes, the additive white Gaussian noise has a power spectral
density of $N_0$. All transmissions in the system have a bandwidth
of $B$ Hz. Assuming that the maximum power MS, relays and BS can
supply are identical, which is $P_0$ joules per second. Thus the SNR
at each receiver is $\rho|h_{i,j}|^2$, where
$\rho:=\frac{P_0}{N_0B}$ is the general SNR without fading.
\par
In addition to SNR, we assume the transmission schemes are further
parameterized by the MS-BS spectral efficiency $R$ bit/s/Hz
attempted by the transmitting terminals\cite{Laneman}. As the
transmission through relay needs two transmission phases, the
spectral efficiency of each phase may be no less than $2R$ to get an
end-to-end $R$ bit/s/Hz spectral efficiency. Fixing $R$ simplifies
the design of the relays as they do not need to remodulate their
transmission using a different signal constellation\cite{patent}.


\par
\subsection{Asymmetric Traffic Model}
To model the asymmetric traffic condition in the system, we
introduce a traffic asymmetry factor $\zeta$, which gives the ratio
of the traffic load $L_{UL}$ in the uplink to the total traffic load
$L_{total}$. $L_{total}$ is the sum of uplink traffic load $L_{UL}$
and the downlink traffic load $L_{DL}$. The traffic load
$L_{total}$, $L_{UL}$ and $L_{DL}$ are defined as the number of
generated bits in a communication. Hence, the $L_{UL}$ and $L_{DL}$
can be obtained by $\zeta$ and $L_{total}$. In cellular systems,
$\zeta$ can
 be obtained from upper layer at BS before relay selection.

\subsection{Weighted Total Energy Consumption Model of Cooperation}
In the scope of energy efficiency estimation methodologies, a
measure usually taken for the comparison of radio transmission
technologies is the energy consumption per bit\cite{parameters}.
However a comparison of the sheer energy consumption of the
transmission schemes in cooperative networks is not suitable,
because certainly MS, relay and BS have different power consumption
challenges. Thus, we use the weighted total energy consumption of
MS, relay and BS in the two-way communication, and allocate
different weights to MS, relay and BS, denoted by $\omega_{M}$,
$\omega_R$ and $\omega_{B}$ , respectively. The weights $\omega_M$,
$\omega_R$ and $\omega_B$ are defined based on criteria of priority
level on power consumption of MS, relay and BS. They can be set
arbitrary values according to different condition. An example can
be, $\omega_M=\omega_R=1$ and $\omega_B=0$. In this case, we give
equal weight to MS and relay, and the energy consumption of BS is
ignored. This is reasonable in user cooperative cellular networks,
as MS and relay are all powered by batteries while BS is always
powered by a fixed line\cite{Cooperation_Book}.
\par
We model only the energy required for radio transmission and not the
energy consumed for receiving. This is reasonable as the radio
transmission is the dominant component of energy consumption for
long range transmissions \cite{GoldSmith}. Thus, the weighted total
energy consumption of the cooperative communication per information
bit can be expressed as
\begin{equation}
\label{eq4}
\begin{split}
E_{coop}&=\frac{1}{2RB}\biggl[ \zeta(\omega_M\cdot
P^{UL}_M+\omega_R\cdot
P^{UL}_R)\\
& \phantom{=\frac{1}{2RB}\biggl[} +(1-\zeta)(\omega_R\cdot
P^{DL}_R+\omega_B\cdot P^{DL}_B) \biggr]
\end{split}.
\end{equation}

\par

\section{Joint Uplink and Downlink Relay Selection}
\label{JUDRS} This section presents a relay selection protocol for a
multi-relay network, called Joint Uplink and Downlink Relay
Selection (JUDRS). We proceed JUDRS as follows.
\par


\textbf{Step1:} MS broadcasts a RTS1 packet to the relays and BS
using a fixed transmission power $P_0$. Each relay hears the RTS1
packet and estimates the gain of the channel between MS and itself,
denoted by $|h_i|^2$. Depending on the channel states, only a subset
$\Gamma$ of the $N$ relays can be chosen as candidate relays,
defined by
\begin{equation}
\label{gamma_set} \Gamma\triangleq \{i\in{S}:P_0\cdot|h_i|^2 \geq
N_0B(2^{2R}-1)=th_1\}.
\end{equation}
In (\ref{gamma_set}), the relay is assumed to be helpful to the
transmission from MS to BS only if $P_0\cdot|h_i|^2 \geq th_1$
\cite{Laneman}. We use $t$ to denote the size of $\Gamma$. BS also
measures the channel gain of the direct link, which is denoted as
$|h_{direct}|^2$.
\par

\textbf{Step2:} The $t$ relays in $\Gamma$ send RTS2 packets to the
BS along with the channel quality indicator (CQI) using power $P_0$.

\textbf{Step3:} BS estimates the channel gain $g_i$ between it and
relay $i$, $i\in\Gamma$. The relay can be helpful to the
transmission between MS and BS only if
\begin{equation}
\label{th_2} P_0\cdot |g_i|^2 \geq
N_0B(2^{2R}-1)\cdot(1-\frac{|h_{direct}|^2}{|h_i|^2})\triangleq
th_2.
\end{equation}
 The relays satisfy (\ref{th_2}) form the candidate relay
set $\Sigma$.
\par
\textbf{Step4:} BS selects the best relay from $\Sigma$ under the
following criteria, and broadcasts the index of the best relay along
with the transmitting power allocated to MS and the relay in UL and
DL transmissions. In order to minimize the total energy consumption,
we optimize the transmitting power of all transmitters in both UL
and DL to the minimum required for successful transmission at an
end-to-end data rate $R$, which are:
\begin{equation}
\label{power}
 P_{M}^{UL}=\frac{th_1}{|h_i|^2},
P_{R}^{UL}=\frac{th_2}{|g_i|^2}, P_{B}^{DL}=\frac{th_1}{|g_i|^2},
P_{R}^{DL}=\frac{th_3}{|h_i|^2},
\end{equation}
where $th_3=N_0B(2^{2R}-1)\cdot(1-\frac{|h_{direct}|^2}{|g_i|^2})$.
 Substituting (\ref{power}) into (\ref{eq4}), the weighted
total energy consumption of MS, relay and BS per information bit can
be expressed as,
\begin{equation}
\label{E_coop}
\begin{split}
E_{coop}
&=\frac{N_0(2^{2R}-1)}{2R}\cdot(\frac{\zeta\cdot\omega_M+(1-\zeta)\cdot\omega_R}{|h_i|^2}\\
&\quad
+\frac{\zeta\cdot\omega_R+(1-\zeta)\cdot\omega_B}{|g_i|^2}-\frac{\omega_R\cdot|h_{direct}|^2}{|h_ig_i|^2})
\end{split}.
\end{equation}
Note that MS can also communicate with BS directly without the help
of any relay if the channel gain of the direct link is strong
enough. Thus, we calculate the weighted energy consumption per bit
required for direct transmission between MS and BS for successful
transmission at data rate $R$,
\begin{equation}
\begin{split}
E_{direct} &=\frac{\zeta\cdot\omega_M\cdot
P_M^{UL}+(1-\zeta)\cdot\omega_B\cdot P_B^{DL}}{R\cdot B}\\
&=\frac{N_0(2^R-1)}{R}\cdot
\frac{\zeta\cdot\omega_M+(1-\zeta)\cdot\omega_B}{|h_{direct}|^2}
\end{split}.
\end{equation}

In this paper, we select the relay which minimizes the weighted
total energy consumption of communication as best relay, which can
be expressed as,
\begin{equation}
i^{\ast} = \arg\min\limits_{i\in\Sigma}\{E_{coop}\}.
\end{equation}
BS compares the $E_{coop}^{i^{\ast}}$ with $E_{direct}$, and choose
the one with the minimum weighted energy consumption. If $E_{direct}
\leq E_{coop}^{i^{\ast}}$, no relay is selected as the direct
transmission is more energy-efficient than the transmission via any
relay.
\par

\textbf{Step5:} MS and BS communicate with each other in uplink and
downlink via the selected relay. If the direct link is selected, MS
and BS communicate directly with each other.
\par
\begin{figure}
\centering
\includegraphics[width=2.9in]{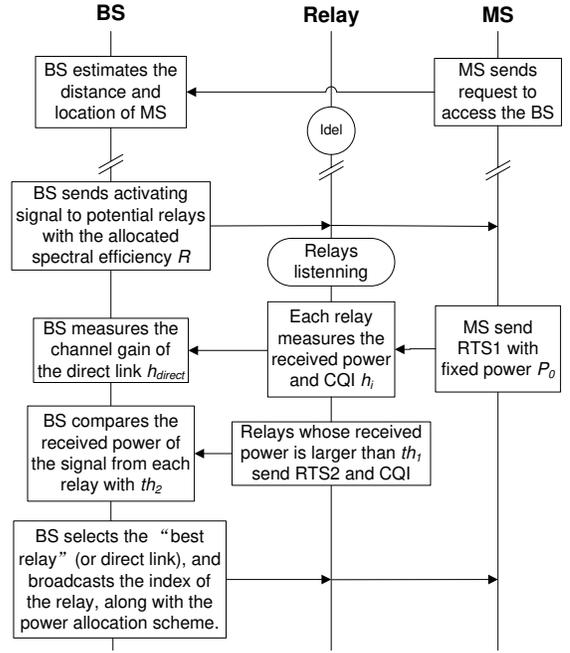}
\caption{The signaling of the JUDRS scheme.} \label{signaling}
\end{figure}

Fig. \ref{signaling} shows the signaling flow of the JUDRS scheme.
It can be seen that the scheme is centralized, most operation are
done at MS, thus it avoid data collision and reduce the power
consumption that MS and relay require to perform relay selection.

\par
\section{Diversity-Multiplexing Tradeoff of JUDRS}
\label{DMT} This section analyzes the diversity-multiplexing
tradeoff (DMT) performance of the proposed relay selection scheme.
We use the definition given in \cite{L_zheng}. A channel is said to
achieve multiplexing gain $r$ and diversity gain $d$ if there exists
a sequence of codes $C(\rho)$ operating at SNR $\rho$ with rate
$R(\rho)$ and resulting outage probability $P_{out}(\rho)$ such
that:
\begin{equation}
\lim\limits_{\rho\rightarrow\infty} \frac{R(\rho)}{\log(\rho)}=r ,
\quad \quad\quad\quad \lim\limits_{\rho\rightarrow\infty} \frac{\log
P_{out}(\rho)}{\log(\rho)}=-d.
\end{equation}
In the following developments, we say $f(\rho)$ is exponentially
equal to $\rho^{v}$, denoted by $f(\rho)\doteq\rho^{v}$, if
\begin{equation}
\lim\limits_{\rho\rightarrow\infty} \frac{\log
f(\rho)}{\log(\rho)}=v.
\end{equation}
We can define $\dot{\leq}$ in a similar fashion.
\par
The main DMT result for JUDRS scheme is given in the following
theorem, where we denote $(\cdot)^{+}=\max\{\cdot,0\}$.
\par
\newtheorem{theorem}{Theorem}
\begin{theorem} The JUDRS scheme achieves the following diversity-multiplexing
tradeoff:
\begin{equation}
d_{JUDRS}(r)=(N+1)\left(1-\frac{2N+1}{N+1}r\right)^{+}.
\end{equation}
\end{theorem}
\par

\begin{IEEEproof}
The outage probability of the JUDRS scheme can be expressed as,
\begin{equation}
P_{out}=\zeta\cdot P_{out}^{UL}+(1-\zeta)\cdot P_{out}^{DL}.
\end{equation}
For the uplink transmission, during the broadcast phase, the mutual
information across the MS-BS channel is:
\begin{equation}
I_{D}=\log(1+\rho|h_{direct}|^2).
\end{equation}
If a retransmission occurs, the combination of the two transmission
forms an equivalent channel between the MS and BS, whose mutual
information is:
\begin{equation}
I_{2-hop}=\frac{1}{2}\log\left(1+\rho(|h_{direct}|^2+|g_{i^*}|^2)\right),
\end{equation}
where $i^*$ denotes the index of the selected relay. Using the law
of total probability, the outage probability of uplink transmission
can be expressed as:
\begin{equation}
\begin{split}
\label{P_out_UL} P_{out}^{UL}&= \sum\limits_{t=1}^{N}Pr\{I_D<R,
I_{2-hop}<R\big| |\Gamma|=t\}\cdot Pr\{|\Gamma|=t\}\\
&\quad +Pr\{|\Gamma|=0\}\cdot Pr\{I_D<R\}\\
&= \biggl[\sum\limits_{t=1}^{N}Pr\{I_{2-hop}<R\big|I_D<R,|\Gamma|=t\}\cdot Pr\{|\Gamma|=t\}\\
&\quad +Pr\{|\Gamma|=0\}\biggr]\cdot Pr\{I_D<R\}
\end{split}.
\end{equation}
The probability that exact $t$ nodes know the
message is given by\cite{ITRS}:
\begin{equation}
\label{gamma=t}
\begin{split}
Pr\{|\Gamma|=t\}&={N\choose t}  \prod_{i \in \Gamma} \exp
\left(-\lambda_{M,i}\frac{2^{2R}-1}{\rho}\right)\\
&\quad \cdot \prod_{i \not\in \Gamma} \left[1-\exp\left(-\lambda_{M,i}\frac{2^{2R}-1}{\rho}\right)\right]\\
&\doteq \rho^{(2r-1)(N-t)}\prod_{i \not\in \Gamma}
\left(\frac{1}{\lambda_{M,i}}\right)
\end{split}.
\end{equation}
The probability that the direct link is in outage is given by,
\begin{equation}
\label{1_hop_fail}
\begin{split}
Pr\{I_D\leq R\} &= Pr\{(1+|h_{direct}|^2)\leq r\log\rho\}\\
&\doteq \rho^{r-1}
\end{split}.
\end{equation}
And the condition probability that the 2-hop channel is in outage
can be calculated as,
\begin{equation}
\label{2_hop_fail}
\begin{split}
&Pr\{I_{2-hop}<R\big|I_D<R,|\Gamma|=t\}\\
&=Pr\{\frac{1}{2}\log(1+\rho(|h_{direct}|^2+|\max\limits_{i\in\Gamma}\{g_i\}|^2))\leq
r\log\rho\\
&\quad\quad\quad \big| 1+\rho |h_{direct}|^2\leq \rho^{r},
\Gamma \}\\
 &\leq Pr\{ |h_{direct}|^2+|\max\limits_{i\in\Gamma}\{g_i\}|^2\leq
\rho^{2r-1}\big||h_{direct}|^2\leq \rho^{r-1},\Gamma\}\\
 &\leq Pr\{|\max\limits_{i\in\Gamma}\{g_i\}|^2\leq\rho^{2r-1}\big|\Gamma\}\\
&\overset{(a)}{\doteq} \rho^{t(2r-1)}
\end{split}.
\end{equation}
Here (a) follows from Lemma 2 in \cite{opportunistic}.
\par
Substituting (\ref{gamma=t}), (\ref{2_hop_fail}) and
(\ref{1_hop_fail}) into (\ref{P_out_UL}), we can get the upper bound
of overall uplink outage probability:
\begin{equation}
\begin{split}
P_{out}^{UL}&\dot{\leq}
\rho^{r-1} \left(\sum\limits_{t=1}^{N}\rho^{t(2r-1)}\rho^{(N-t)(2r-1)}+\rho^{N(2r-1)}\right)\\
&\doteq \rho^{(2N+1)r-(N+1)}
\end{split}.
\end{equation}
The same methodology can be applied to downlink transmission. Thus
the DMT of the JUDRS scheme is given by
\begin{equation}
d_{JUDRS}(r)=(N+1)\left(1-\frac{2N+1}{N+1}r\right)^{+}.
\end{equation}
\end{IEEEproof}
\par
The DMT analysis corroborates the merits of the JUDRS scheme that it
can achieve full spatial diversity in the number of the cooperating
nodes, not just the number of decoding relays.

\begin{figure*}
\centering \subfigure[$R=3$bit/s/Hz, $d_{M,B}=450$m]{
\includegraphics[width=2.8in]{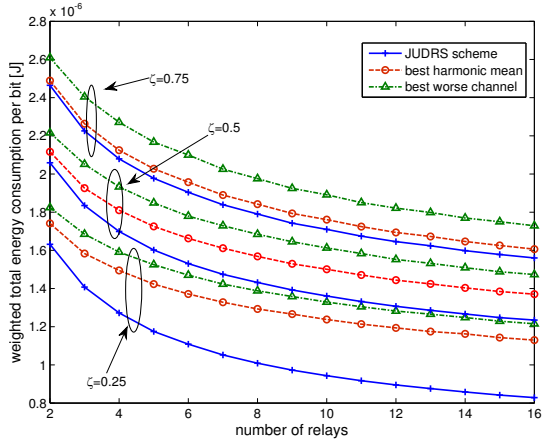}
\label{D450}} \hfil \subfigure[$R=1$bit/s/Hz, $d_{M,B}=1200$m]{
\includegraphics[width=2.8in]{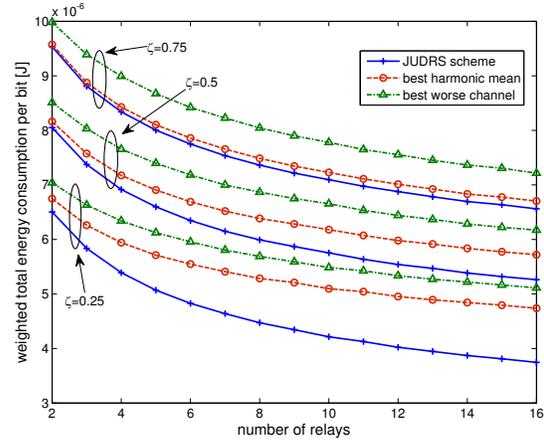}
\label{D1200}} \caption{Total transmission energy of MS and relay
per bit in two topologies} \label{two_pic}
\end{figure*}

\section{Simulation Results}
\label{numerical results} 
\begin{figure}
\centering
\includegraphics[width=2.8in]{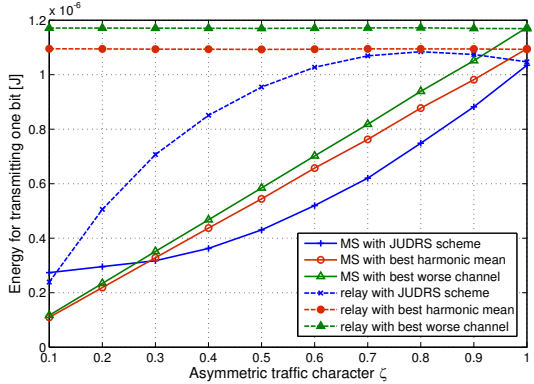}
\caption{The energy consumption of MS and relay per bit for
different traffic conditions, with 8 active relays and MS-BS
distance of 450m, $R=3$bit/s/Hz.} \label{change_k}
\end{figure}

In this section, we present the
performance results of the JUDRS protocol. For comparison, we also
simulate the performance of some existing relay selection schemes.
In \cite{best_worse}, the best worse channel selection is
introduced, in which the relay whose worse channel,
$\min\{|h_i|,|g_i|\}$, is the best is selected. In
\cite{opportunistic}, the best harmonic mean selection is proposed,
in which the relay selection function is chosen as the harmonic mean
of the two channel's magnitudes: $(|h_i|^{-2}+|g_i|^{-2})^{-1}$. The
relay with the largest harmonic mean cooperates. We extend the best
worse channel selection and the best harmonic harmonic mean
selection to both uplink and downlink straightforward. The
parameters used in the simulation is shown in Table \ref{table_1},
which are mainly taken from \cite{parameters}, where $f_c$ is the
frequency of the central carrier. Throughout the simulations, we use
$\omega_M=\omega_R=1$ and $\omega_B=0$. The simulation results are
shown in Fig.~\ref{two_pic} and Fig. \ref{change_k}.


\begin{table} [h]
\centering \caption{\label{table_1}Simulation Parameters}
\begin{tabular}[t]{c|c|c|c}
\hline
$P_0$ & 24dBm & $G_T\cdot G_R$ & 5dBi\\
$B$ & 180KHz  & $f_c$ & 2.5GHz \\
$N_0$ & -171dBm/Hz & $p$ & 3.76 \\
\hline
\end{tabular}
\end{table}
\par
In Fig. \ref{D450} and Fig. \ref{D1200}, the total transmission
energy of MS and relay per bit is plotted as a function of the
number of relays for two scenarios: a direct link between MS and BS
exists (see Fig. \ref{D450}, MS-BS distance of 450m) and a direct
link does not exists (see Fig. \ref{D1200}, MS-BS distance of 1200m,
$|h_{direct}|\ll \min\{|h_i|,|g_i|\}$). These figures indicate that
the proposed JUDRS scheme can save the total transmission energy
efficiently than the best worse channel selection and the best
harmonic selection in both cases with different traffic load
conditions. It is also shown in the figures that the energy consumed
for transmitting one bit decreases with the active relay number,
which means that it is helpful to have a larger number of relays.
\par
Fig. \ref{change_k} compares the energy consumption of MS and relay
for transmitting one bit under different traffic conditions when 8
relays are active. It is shown that both MS and relay consume less
energy using the JUDRS scheme than the best harmonic mean selection
and best worse channel selection when $\zeta > 0.3$. Fig.
\ref{change_k} also illustrates that the energy saving achieved by
the proposed scheme grows with the decrease of the asymmetric
traffic factor $\zeta$. It gives the advantage of the JUDRS scheme
that it can adaptively adjust the transmit energy to the minimum
according to the traffic condition in the system. As the uplink
traffic load is always less than that of the downlink, we can always
expect a larger energy-saving using the JUDRS scheme. Fig.
\ref{change_k} also indicates the energy-saving achieved by the
JUDRS scheme mainly comes from the energy saving of the relay,
especially when the uplink traffic load is less than the downlink
traffic load.

\section{Conclusion}
\label{conclusion} This paper presents an energy-efficient relaying
scheme through selection of mobile relays. We propose Joint Uplink
and Downlink Relay Selection (JUDRS) scheme, in which relay is
selected jointly for uplink and downlink, and the weighted total
energy consumption is minimized. Power allocation policies are also
designed for communication nodes to improve the overall energy
efficiency. The diversity-and-multiplexing tradeoff result
demonstrates that the JUDRS scheme can achieve full spatial
diversity in the number of cooperating terminals. Numerical results
further confirm that the proposed scheme can improve the
energy-efficiency of MS and relay compared with previous best
harmonic mean selection and best worse channel selection.






%

\end{document}